\documentclass[preprint2]{aastex}
\usepackage{aastexug}

\def\lesssim{\mathrel{\hbox{\rlap{\hbox{\lower4pt\hbox{$\sim$}}}\hbox{$<$}}}}
\def\gtrsim{\mathrel{\hbox{\rlap{\hbox{\lower4pt\hbox{$\sim$}}}\hbox{$>$}}}}

\newcommand{\thetavec}{\mbox{\boldmath $\theta$}}
\newcommand{\zetavec}{\mbox{\boldmath $\zeta$}}

\newcommand{\uvec}{\mbox{\boldmath $u$}}

\begin{document}

\title{Probing the Spatial Distribution of Extrasolar Planets with 
	   Gravitational Microlensing} 

\author{Cheongho Han}
\affil{Department of Physics, Institute for Basic Science Researches, 
Chungbuk National University, Chongju 361-763, Korea}
\email{cheongho@astroph.chungbuk.ac.kr} 

\author{Young Woon Kang}
\affil{Department of Earth Sciences, Sejong University,
       Seoul 143-747, Korea}
\email{kangyw@sejong.ac.kr}

\begin{abstract}
To search for extrasolar planets, current microlensing follow-up 
experiments are monitoring events caused by stellar-mass lenses hoping 
to detect the planet's signature of the short-duration perturbation in 
the smooth lensing light curve of the primary.  According to this strategy, 
however, it is possible to detect only planets located within a narrow 
region of separations from central stars.  As a result, even if a large 
sample of planets are detected and the separations from their central 
stars are determined, it will be difficult to draw meaningful result 
about the spatial distribution of planets.  An additional channel of 
microlensing planet detection is provided if the monitoring frequency 
of survey experiments is dramatically increased.  From high-frequency 
monitoring experiments, such as the recently proposed GEST mission, one 
can detect two additional populations of planets, which are free-floating 
planets and bound planets with wide orbits around central stars.  In this 
paper, we investigate the lensing properties of events caused by wide-orbit 
planets and find that the light curves of a significant fraction of these 
events will exhibit signatures of central stars, enabling one to distinguish 
them from those caused by free-floating planets.  Due to the large 
primary/planet mass ratio, the effect of the central star endures to 
considerable separations.  We find that for a Jupiter-mass planet the 
signatures of the central star can be detected with fractional deviations 
of $\geq 5\%$ from the best-fitting single-lens light curves for 
$\gtrsim 80\%$ of events caused by bound planets with separations 
$\lesssim 10$ AU and the probability is still substantial for planets 
with separations up to $\sim 20$ AU.  Therefore, detecting a large sample 
of these events will provide useful information about the distribution 
of extrasolar planets around their central stars.  Proper estimation of 
the probability of distinguishing events caused by wide-orbit planets 
from those caused by free-floating planets will be important for the 
correct determination of the frequency of free-floating planets, whose 
microlensing sample will be contaminated by wide-orbits planets.
\end{abstract}

\section{Introduction}
A Galactic microlensing event occurs when a compact massive object
(lens) approaches very close to the observer's line of sight toward a 
background star (source).  Due to lensing, the source star image is 
split into two with different fluxes from that of the unlensed source.
The locations and magnifications of the individual images are
\begin{equation}
\thetavec_\pm = {1\over 2}\left( \uvec \pm \sqrt{u^2+4}{\uvec\over u}
\right) \theta_{\rm E},
\end{equation}
and 
\begin{equation}
A_\pm = {u^2+1\over 2u\sqrt{u^2+4}} \pm {1\over 2},
\end{equation}
where $\uvec$ is the projected lens-source separation vector normalized by 
the Einstein ring radius $\theta_{\rm E}$.  The Einstein ring represents 
the effective lensing region around the lens within which the combined 
source star flux is magnified greater than $3/\sqrt{5}\sim 1.34$.  For a 
typical Galactic bulge event with a lens and a source located at $D_{ol}=6$ 
kpc and $D_{os}=8\ {\rm kpc}$, respectively, the Einstein ring has a radius 
of 
\begin{equation}
\theta_{\rm E} \sim 0.32\ {\rm mas}\ 
\left( {m\over 0.3 M_\odot} \right)^{1/2},
\end{equation}
where $m$ is the lens mass.  The angular size of $\theta_{\rm E}$ corresponds 
to the physical distance at the lens location of 
\begin{equation}
r_{\rm E} = D_{ol}\theta_{\rm E} \sim 2\ {\rm AU}\ 
\left( {m\over 0.3 M_\odot}\right)^{1/2}.
\end{equation}
For Galactic events, the separation between the two images is
$\left\vert\thetavec_+-\thetavec_-\right\vert=\sqrt{u^2+4}\theta_{\rm E}
\sim 2\theta_{\rm E} \sim 0.6\ {\rm mas}$, which is too small for the images 
to be resolved.  However, the flux of the combined image varies with time 
due to the relative motion of the observer, lens, and source, and thus 
lensing events can be identified from the variation of source star fluxes.
The light curve of a lensing event is represented by 
\begin{equation}
A = A_+ + A_- = {u^2+2 \over u(u^2+4)^{1/2}}.
\end{equation}
Light variation caused by lensing can be distinguished from other types of 
variations due to the smooth, symmetric, and non-repeating characteristics 
of the lensing light curves.  The duration of a lensing 
event is characterized by the Einstein time scale $t_{\rm E}$, which 
represents the time required for the source to transit $\theta_{\rm E}$.  
For a typical lens-source proper motion of $\mu_{rel} \sim 25\ {\rm km}
\ {\rm s}^{-1}\ {\rm kpc}^{-1}$, the Einstein time scale has a value of 
\begin{equation}
t_{\rm E} = {\theta_{\rm E} \over \mu_{rel}} 
\sim 20\ {\rm days} 
\left( {m\over 0.3\ M_\odot} \right)^{1/2}.
\end{equation}

If an event is caused by a lens having a planetary-mass companion and 
the position of the companion happens to be near the path of one of the 
two images created by the primary lens, the planet will perturb the light 
from the nearby image, causing deviation in the lensing light curve of the 
primary \citep{mao1991, gould1992}.  For a planet with a mass ratio $q$ 
to the central star, the deviation lasts for a duration of $t_{{\rm E},p} 
\sim \sqrt{q}t_{\rm E}$, which corresponds to $\sim 1$ day for a Jupiter-mass 
planet.  Due to the short duration, it is difficult to detect the 
planet-induced perturbations from the current survey-type experiments, 
which have typical monitoring frequencies of 1--2/night.  To increase the 
monitoring frequency, current experiments are employing early warning systems 
\citep{alcock1996, afonso2001, bond2001, udalski1994} to issue alerts of 
ongoing events detected in the early stage of lensing magnification and 
follow-up observation programs \citep{alcock1997, rhie1999, albrow1998} 
to intensively monitor the alerted events.  Once the deviation is detected 
and analyzed, one can determine the mass ratio and the projected separation 
(normalized by $\theta_{\rm E}$) between the planet and 
central star, $d$ \citep{gaudi1997}.

One important drawback of the current microlensing planet search strategy 
is that one can detect only planets located within a narrow region of 
separations from central stars.  This is because the images produced by 
the central star are located close to the Einstein ring during the event, 
and thus only planets located near the ring can effectively perturb the 
images (see more details in \S\ 2).  As a result, even if a large sample 
of planets are detected and their separations are determined, no meaningful 
result can be drawn about the spatial distribution of planets around central 
stars.

A Second channel for the discovery of planets via microlensing is provided 
by planets in wider orbits.  When the orbit of a planet is wide, two types 
of events are expected.  One is ``repeating'' events in which the source 
trajectory passes close to both the planet and primary star and the other 
is ``isolated'' events in which the trajectory passes close to the planet 
only.  Planets via the channel of repeating events can be detected by the 
current-type follow-up observations [See the extensive works of Di Stefano 
\& Scalzo (1999a,1999b) on this channel of planet detections].  Detecting 
planets via the channel of isolated events, on the other hand, is difficult 
by the current experiments because these events occur without any warning 
and last for very short period of time.  However, if the monitoring frequency 
is dramatically increased, it will be possible to detect a large sample of 
wide-orbit planets through this channel.  Recently, such a high-frequency 
survey experiment (GEST) was proposed to NASA by \citet{bennett2002}.  The 
GEST mission is designed to continuously monitor $\sim 10^8$ Galactic bulge 
main-sequence stars with a frequency of several times per hour by using a 
1--2 m aperture space telescope.  Another population of planets that can 
be detected by the high-frequency lensing survey is free-floating planets.

In this paper, we investigate the lensing properties of events caused by 
wide-orbit planets that will be detected by future high-frequency lensing 
experiments.  Unlike the negligible effect of the wide-orbit planet on the 
lensing behavior of the primary, the effect of the primary on the lensing 
behavior of the planet may be important even at very large separations due 
to the large primary/planet mass ratio.  If this is so, the signature of 
the central stars might be noticed for a significant fraction of events, 
enabling one to distinguish them from events caused by free-floating 
planets.

The paper is composed as follows.  In \S\ 2, we discuss the basics of 
planetary microlensing that are required to describe the lensing behavior 
of events caused by wide-orbit planets.  In \S\ 3, we investigate the 
properties of wide-orbit planetary lensing in detail.  In \S\ 4, we 
estimate the probability of distinguishing events produced by wide-orbit 
planets from those caused by free-floating planets.  In \S\ 5, we 
summarize the results and discuss the implications 
of the results.

\section{Basics of Planetary Lensing}

The lensing behavior of a system having a planetary-mass companion is 
described by the formalism of binary lensing with a very low mass-ratio 
companion.  If a source star located at $\zeta=\xi+i\eta$ in complex 
notations is lensed by two point-mass lenses with the individual locations 
of $z_{L,1}=x_{L,1}+i y_{L,1}$ and $z_{L,2}=x_{L,2}+i y_{L,2}$ and the 
mass fractions of $m_1$ and $m_2$, respectively, the locations of the 
resulting images $z=x+iy$ are obtained by solving the lens equation, 
which is represented by
\begin{equation}
\zeta = z + {m_1 \over \bar{z}_{L,1}-\bar{z}} + {m_2 \over 
\bar{z}_{L,2}-\bar{z}},
\end{equation}
where $\bar{z}$ denotes the complex conjugate of $z$ and all lengths 
are normalized by the Einstein ring radius corresponding to the total 
mass of the binary (combined Einstein ring radius $\theta_{\rm E}$).
Since the lens equation describes a mapping from the lens plane to the 
source plane, finding image positions ($x,y$) for a given source position 
($\xi,\eta$) requires inverting the lens equation.  Although the lens 
equation for a binary lens system cannot be algebraically inverted due 
to its nonlinearity, it can be expressed as a fifth-order polynomial 
in $z$ and the image positions can be obtained by numerically solving the 
polynomial \citep{witt1990}.  Since the lensing process conserves the 
source star surface brightness, the magnification of each image equals to 
the area ratio between the image and the unlensed source and mathematically 
it is obtained by computing the Jacobian of the mapping equation evaluated 
at the image position;
\begin{equation}
A_i = \left\vert \left( 1-{\partial\zeta\over\partial\bar{z}}
{\overline{\partial\zeta}\over\partial\bar{z}} \right)^{-1} \right\vert.
\end{equation}
Then, the total magnification is given by the sum of the magnifications 
of the individual images, i.e.\ $A=\sum_i A_i$.

Due to the very small mass ratio of the planet to the primary, the 
planetary lensing behavior is well described by that of a single lens 
event for most of the event duration.  However, noticeable deviations 
can occur if the planet is located close to one of the images produced 
by the primary lens.  The region around the image-perturbing planet's 
location in the lens plane corresponds to the region around caustics 
in the source plane.  Therefore, for the source-plane point of view, 
noticeable deviations occur when the source approaches the region around 
caustics.  The caustics are the main new features of binary lensing and 
refer to the set of source positions at which the magnification of a point 
source becomes infinity.  For the case of a wide-orbit planet, the caustic 
is located along the primary-planet axis and its location on the axis is 
approximated by
\begin{equation}
x_c \sim x_p - {1\over x_p},
\end{equation}
where $x_p$ is the position of the planet (in units of $\theta_{\rm E}$) 
with respect to the primary which is located at the origin \citep{griest1998}.
Then, caustics are located within the Einstein ring when the planetary 
separation is in the range of $0.6\lesssim x_p \lesssim 1.6$, which is so 
called the `lensing zone' \citep{gould1992, wambsganss1997}.  In addition,
the size of the caustic, and thus the probability of detecting planet-induced 
deviations, is maximized when the planet is in the lensing zone.  Under the 
current planet-search strategy of monitoring events caused by stellar-mass 
lenses, therefore, only planets within the lensing zone can be effectively 
detected.

\section{Lensing by Wide-orbit Planet}

As the separation between the planet and central star increases, the 
caustic shrinks rapidly.  If the separation is significantly larger than 
the Einstein ring radius, both lens components behave as if they are two 
independent single lenses.  Due to the deflection of light produced by 
the presence of the companion, however, the position of each lens is 
effectively shifted towards the companion \citep{distefano1996}.  The 
effective positions of the individual lenses are given by
\begin{equation}
\tilde{x}_{L,i} = x_{L,i} - {m_j/m_i \over \ell/\theta_{{\rm E},i} } 
\ {\rm sign} (x_{L,i}-x_{L,j}),
\end{equation}
where $\theta_{{\rm E},i}$ represents the Einstein ring radius of each 
lens, the subscripts `$i$' and `$j$' are used to denote one lens 
component and its companion, respectively, and the term $-{\rm sign} 
(x_{L,i}-x_{L,j})$ implies that the shift is towards the direction of the 
companion.  Then, the amount of the positional shift of the central star 
due to the planet is $\delta_\ast=\vert\tilde{x}_\ast-x_\ast\vert\sim q/d$, 
which is negligible due to the combination of small $q$ and large 
$d$.  On the other hand, the amount of the shift of the planet's position 
due to its central star, $\delta_p = \vert \tilde{x}_p-x_p \vert \sim 1/d$, 
is not negligible because it does not depend on $q$.  Note that the 
effective position of the wide-orbit planet is 
\begin{equation}
\tilde{x}_p \sim x_p - {1 \over d} = x_p - {1 \over x_p},
\end{equation}
implying that it corresponds to the position of the caustic [cf.\ eq.\ (9)]. 
Since significant deviations can occur only if the source trajectory passes  
close to caustics, although most part of the light curve of the event 
caused by a wide-orbit planet is approximated by that of a single-lens 
event produced by the planet at its effective position, its central part 
can be distorted due to the effect of the central star.

Another thing to be taken into consideration in describing the lensing 
behavior of events caused by wide-orbit planets is the effect of extended 
source size.  The finite-source effect for these events is important 
because the Einstein ring radius of the planet, $\theta_{{\rm E},p}=\sqrt{q}
\theta_{\rm E}$, is very small, and thus the lensed star can no longer be 
approximated by a point source \citep{bennett1996}.  For events caused a 
Jupiter-mass planet, for example, the radii of source stars correspond to 
$\rho_\ast\sim 3.3\%$, 10\%, and 43\% of $\theta_{{\rm E},p}$ for events 
involved with main-sequence ($R_\ast\sim 1\ R_\odot$), turn-off 
($R_\ast\sim 3\ R_\odot$), and clump giant ($R_\ast\sim 13\ R_\odot$) source 
stars, respectively.  The lensing magnification affected by the finite-source 
effect is given by the intensity-weighted magnification averaged over the 
source flux, i.e.\
\begin{equation}
A_{fs}={\int\int I(\zetavec') A(\zetavec+\zetavec') d\Sigma_\ast
\over 
\int\int I(\zetavec') d\Sigma_\ast },
\end{equation}
where $\zetavec$ is the vector notation of the center of the source, 
$\zetavec'$ is the displacement vector of a point on the source star 
surface with respect to the source star's center, and the notation 
$\int\int \cdot\cdot\cdot\ d\Sigma_\ast$ represents the surface integral 
over the source star surface.  Under the assumption that the source has 
a uniform surface brightness profile, the computation of the magnification 
can be reduced from a two-dimensional to a one-dimensional integral by 
using Green's theorem \citep{gould1997, dominik1998}.

To investigate the lensing properties of events caused by wide-orbit 
planets in more detail, we construct maps of fractional magnification 
excesses.  The magnification excess is defined by 
\begin{equation}
\epsilon = {\left\vert A - A_0 \right\vert \over A_0},
\end{equation}
where $A$ is the exact magnification of the wide-orbit planet and $A_0$ 
is the magnification of the single lens approximation, i.e.\ that of a 
free-floating planet.  The map provides an overview of the anomaly 
patterns, not testing all light curves resulting from a large number of 
source trajectories.

In Figure 1, we present the constructed excess maps of three example lens 
systems having planets with different separations from the central star 
of $d=5$, 7, and 9, respectively.  For the lens, we assume the Einstein 
ring radius is $r_{\rm E}\sim 2$ AU by adopting that of a typical Galactic 
bulge event, and thus the separations correspond to the physical distances 
of $\ell\sim 10$ AU, 14 AU, and 18 AU, respectively.  The mass ratio of the 
tested planet to that of the primary is $q=0.003$, which corresponds to that 
of a Jupiter-mass planet around a low-mass main-sequence star with a mass 
of $0.3\ M_\odot$.  Since the most common stars to be monitored by the GEST 
mission will be main-sequence stars, the maps are constructed assuming that 
the source star has a radius of $R_\ast=1\ R_\odot$.  Each map is centered 
at the effective position of the planet (marked by `+') and the true position 
of the planet is marked by a filled dot.  All lengths are normalized by 
$\theta_{{\rm E},p}$.  From the maps, one finds that, as expected, the 
effective planet position matches very well the center of the caustic induced 
by the central star.  One also finds that the size of significant deviation 
regions (compared to the size of the Einstein ring of the planet) is not 
negligible even for planets located at large separations.  In Figure 2, we 
present several example light curves of events resulting from the source 
trajectories marked in Fig.\ 1.

To see the variation of anomaly pattern depending on the planet's mass 
ratio, we also construct maps of planets with different mass ratios and 
present them in Figure 3.  For these maps, the planet-primary separation 
is fixed as $d=7$ ($\ell=14$ AU).  The tested planets have mass ratios of 
$q=0.003$, 0.001, and 0.0001, which correspond to a Jupiter-, Saturn-, 
and $10\ M_\oplus$-mass planet around a $0.3\ M_\odot$ star, respectively.
The dot on the upper right corner of each left panel represents source size 
relative to the Einstein ring of the planet.  We note that since all lengths 
are scaled by  $\theta_{{\rm E},p}$, which is proportional to $\sqrt{q}$, 
the source size appears to be different although the actual size is the 
same. From the maps, we find that finite-source effect does not seriously 
affect the anomaly patterns of events caused by giant planets with 
$q\gtrsim 10^{-3}$.  For planets in this mass regime, the maps have 
similar patterns because both the size of the caustic and the planet's 
Einstein ring radius have the same dependency on the mass ratio, i.e.\ 
$\propto \sqrt{q}$, and thus they decrease by the same scale as $q$ 
decreases.  This can be seen from the similarity between the maps of 
Jupiter- and Saturn-mass planets.  As the mass ratio further decreases, 
however, the finite-source effect becomes important. We find that for 
the case of planets with $q\lesssim 10^{-4}$, the source size becomes 
bigger than the size of the caustic even for main-sequence source stars 
and the signatures of the central star is seriously washed out by the 
finite-source effect.

\section{Detection Probability}
By using the excess map, we then estimate the probability of distinguishing 
events caused by wide-orbit planets from those produced by free-floating 
planets.  For this estimation, we assume that events with $A\geq A_{th}$ 
are continuously\footnote{Since the lensing experiment planned by the GEST 
mission will be carried out in space, events will be monitored continuously.} 
monitored with a frequency of 3 times per hours.  Here $A_{th}$ is the 
threshold magnification, which is required for event identification.  For 
each planet separation, we produce 900 light curves resulting from source 
trajectories with orientations relative to the planet-primary axis and impact 
parameters to the planet, which are randomly selected in the range of 
$0\leq \alpha \leq 2\pi$ and $0\leq u_p \leq u_{th}$, respectively.  Here 
the quantity $u_p$ represents the impact parameters of the source trajectory 
to the planet normalized by $\theta_{{\rm E},p}$ and $u_{th}$ is the impact 
parameter corresponding to $A_{th}$.  With the excess map, the light curves 
are produced by the one-dimensional cut through the map.  Since the map is 
constructed by considering the finite-source effect, the light curves produced 
in this way automatically incorporate the effect.  To be identified as a bound 
planet, it is assumed that the light curve should have at least one data point 
with deviations greater than a threshold value of $\epsilon_{th}=5\%$.  We note 
that the photometric precision of the GEST mission will be $\lesssim 1\%$ 
even for main-sequence source stars, and thus the adopted detection threshold 
is very conservative choice.  With this detection criteria, the probability 
is determined as the ratio of the number of events with noticeable deviations 
out of the total number of tested events.

In Figure 3, we present the resulting probabilities as a function of 
planetary separation for planets with different mass ratios, which are 
distinguished by different line colors.  The sets of curves with different 
line types are the probabilities for different threshold magnifications.
When the threshold is fixed, the probability becomes smaller as $q$ decreases 
due to the combination of shorter duration of perturbations and larger 
effect of extended sources.  For a given planet, the probability increases 
as the threshold magnification increases.  This is because the deviation 
region is confined to the central region around the effective planet position 
and thus the chances to detect deviations are greater for higher magnification 
events.  From the figure, we find that the signatures of central stars can 
be detected with significant probabilities ($\gtrsim 60\%$) for planets 
with separations $\ell\lesssim 10$ AU, which roughly corresponds to the 
distance of the Saturn from the Sun.  Although the probability drops 
rapidly for planets with $q\lesssim 10^{-4}$, the probability is still 
substantial for giant planets with separations up to $\sim 20$ AU, 
corresponding to the separation between the Uranus and the Sun.

\section{Summary and Conclusion}
Under the current microlensing planet search strategy of monitoring events 
caused by stellar-mass lenses, only planets located within a narrow region 
of separations from central stars can be effectively detected.  However, 
with the dramatic increase of the monitoring frequency, two additional 
populations of free-floating and wide-orbit planets can be detected.  
We investigated the lensing properties of events caused by wide-orbit planets 
and found that a significant fraction of these events could be distinguished 
from those caused by free-floating planets.  We determined that even with 
moderate detection criteria the probability to detect signatures of central 
stars for events caused by wide-orbit planets would be $\gtrsim 80\%$ for 
giant planets with  separations $\lesssim 10$ AU and the probability is still 
substantial for planets with separations up to $\sim 20$ AU.  Detecting a 
large sample of these events will be important because they can provide 
useful information about the spatial distribution of extrasolar planets, 
which cannot be obtained by the sample to be acquired under the current 
microlensing planet search strategy or from other planet search methods such 
as the radial velocity and transit methods.  In addition, proper estimation 
of the probability of distinguishing events caused by the two populations of 
planets will be important for the correct determination of the frequency of 
free-floating planets, whose microlensing sample will be contaminated by 
wide-orbits planets.

\bigskip
We would like to thank J.\ H.\ An for kindly providing his code designed 
for efficient computation of lensing magnifications of events affected by 
finite-source effect.  This work was supported by the Astrophysical 
Research Center for the Structure and Evolution of the Cosmos (ARCSEC) 
of Korea Science \& Engineering Foundation (KOSEF) through Science 
Research Program (SRC) program.

{}

\begin{figure}[ht]
\epsscale{1.3}
\centerline{\plotone{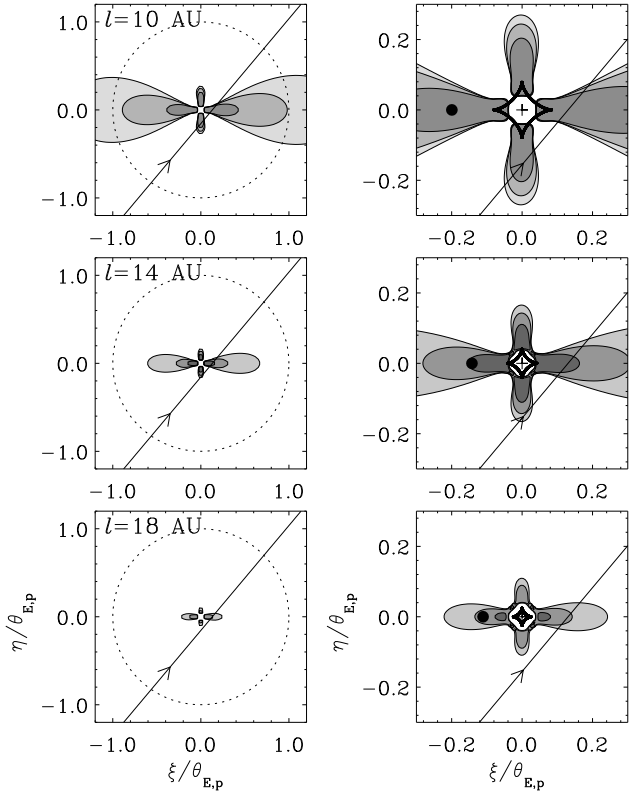}}
\caption{
Contour maps of fractional magnification excesses, $\epsilon$, of planets
with wide separations from central stars.  The left-side panels show 
the maps for planets with projected separation from central stars of 
$l=10$ AU, 14 AU, and 18 AU, respectively, and the right-side panels show 
the enlargements of the maps in the central region of the corresponding 
left-side maps.  For all three cases, planets have a common mass ratio to 
the central star of $q=0.003$.  The locations are set so that the effective 
position of the planet (marked by `+') is at the origin and all lengths are 
normalized by the Einstein ring radius of the planet, $\theta_{{\rm E},p}$.  
The dotted circle on each of the left-side panel represents the Einstein 
ring of the planet.  The closed figure on each of the right panels (drawn 
by the thick solid line) is the caustics.  Both the planet and the central 
star are located on the $\xi$ axis and the central star is on the right.  
The position of the true planet position is marked by a filled dot.  The 
physical separation between the planet and the central star, $\ell$, is set 
by assuming $r_{\rm E}=2$ AU.  The source star is assumed to have a radius 
of $R_\ast=1\ R_\odot$.  Both grey-scales and contours are used to represent 
the regions of significant deviations with $\epsilon\geq 3\%$, 5\%, and 
10\%, respectively.  The straight lines are the source trajectories of 
events whose resulting light curves are presented in Fig.\ 2.
}
\end{figure}

\begin{figure}[ht]
\epsscale{1.1}
\centerline{\plotone{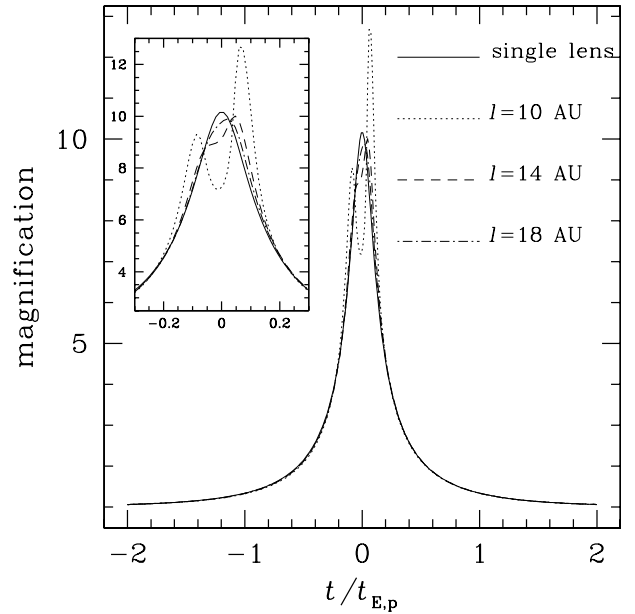}}
\vskip-0.5cm
\caption{
Light curves of events caused by wide-orbit planets with various separations
from central stars.  The source trajectories responsible for the events 
are marked in the corresponding maps in Fig.\ 1.  The time is normalized 
by the Einstein time scale of the planet, $t_{{\rm E},p}$.  The inset shows 
the part of the light curves near the peaks.
}
\end{figure}

\begin{figure}[ht]
\epsscale{1.3}
\centerline{\plotone{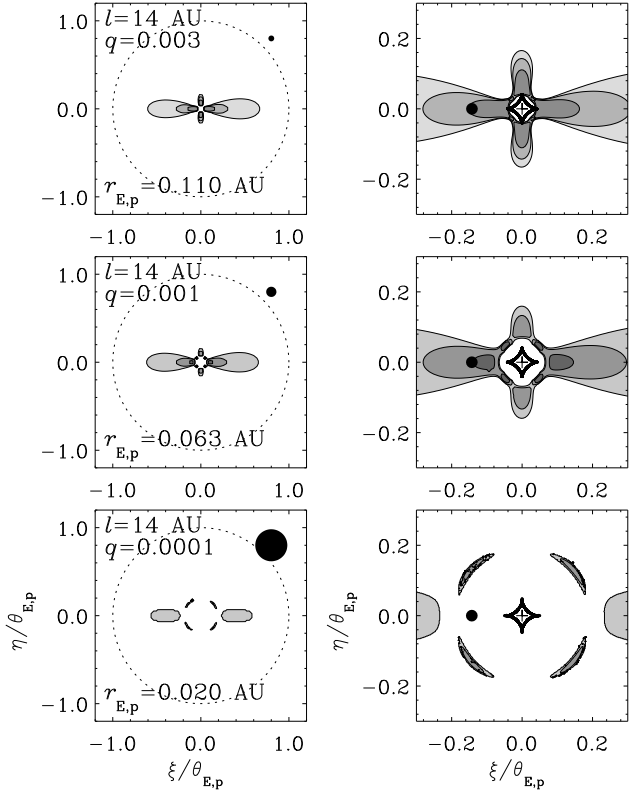}}
\caption{
Excess magnification maps of planets with different mass ratios.  The 
planet-primary separation is fixed as $d=7$ ($\ell=14$ AU) and the tested 
planets have mass ratios of $q=0.003$, 0.001, and 0.0001, which correspond 
to a Jupiter-, Saturn-, and $10\ M_\oplus$-mass planet around a $0.3\ M_\odot$ 
star, respectively.  The filled circle on the right upper corner of each left panel 
represents the source size, which is assumed to be $R_\ast=1\ R_\odot$.
Notations and scales of lengths are similar to those in Fig.\ 1.  Note 
that since all lengths are scaled by $\theta_{{\rm E},p}$, the sizes of the 
individual Einstein rings appear to be the same, although the actual size is 
proportional to $\sqrt{q}$, and the source sizes appear to be different, 
although they are the same.
}
\end{figure}

\begin{figure}[ht]
\epsscale{1.1}
\centerline{\plotone{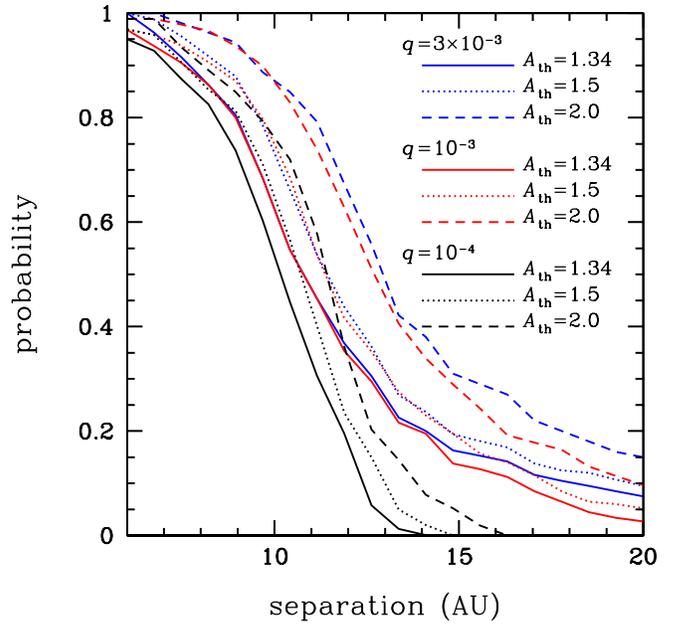}}
\vskip-0.5cm
\caption{
The probability of distinguishing events caused by wide-orbit planets from 
those produced by free-floating planets.  For the observation condition and 
selection criteria used for the identification of central stars, see the 
related text in \S\ 4.
}
\end{figure}

\end{document}